\documentstyle[prl,aps,multicol,epsfig]{revtex}
\begin{document}
\title{The magnetic field dependence of the electronic specific heat of $Y_{0.8}Ca_{0.2}Ba_2Cu_3O_{6+x}$}

\author{J. L. Luo$^1$ ,J. W. Loram$^2$, T. Xiang$^3$, J. R. Cooper$^2$, and J. L. Tallon$^4$}

\address{$^1$Institute of Physics and Center for Condensed Matter
Physics, Chinese Academy of Sciences, P.O. Box 2711, Beijing 100080, The People's Republic of China}

\address{$^2$IRC in Superconductivity, University of Cambridge,
Madingley Road, Cambridge CB3 0HE, United Kingdom }

\address{$^3$Institute of Theoretical Physics, Chinese Academy of Sciences,
P.O.Box 2735, Beijing 100080, The People's Republic of China}

\address{$^4$ NewZealand IRL, P.O. box 31310, Lower Hutt, New Zealand}

\date{\today}
\maketitle
\begin{abstract}

We have measured the electronic specific heat of $Y_{0.8}Ca_{0.2}Ba_2Cu_3O_{7-\delta}$ using a high-resolution
differential technique from liquid helium temperature to room temperature in an applied magnetic field up to 13T.
The field dependence of the electronic specific heat at low temperatures in the superconducting state behaves
differently in the overdoped and underdoped regimes, varing as  $\sqrt{H}$ in the overdoped regime but as $H$ in
the underdoped regime. An entropy loss is observed in the normal state in optimal and underdoped samples, which
can be attributed to the opening of a normal state psuedogap. From the temperature and field dependences of the
free energy, the temperature dependence of the upper critical field $H_{c2}$ is determined. For the overdoped
sample ($x=0.79$), we find $H_{c2}$ to have a negative curvature.

\end{abstract}

\pacs{PACS numbers: 74.25.Bt, 74.60.Ec}

\begin{multicols}{2}

\section{Introduction}
Measurements of electronic specific heat have played an important role in clarifying the nature of superconducting
pairing symmetry\cite{Moler94,Momono94,Wright99,Junod97,Revaz98}, normal state
pseudogap\cite{Loram93,Loram97,Loram00}, quantum criticality\cite{Tallon00}, and many other fundamental properties
of high-$T_c$ superconductors\cite{Junod90,Junod96}. While the mechanism of high-temperature superconductivity is
not understood, there is an emerging consensus that the superconducting state itself is not particularly exotic,
in the sense that it consists of a BCS-like pair state with well-defined quasiparticle excitations. Specific heat
experiments are insensitive to the phase of the order parameter, but can provide bulk information on the behavior
of low energy density of states. It has been confirmed that at very low temperatures below $T_c$ the zero field
specific heat varies as $T^2$ and the field dependent specific heat scales as
$\sqrt{H}$,\cite{Moler94,Wright99,Revaz98} consistent with the d-wave pairing scenario of high-$T_c$
superconductivity. Revaz et.al \cite{Revaz98} also showed that the field dependent term obeys a scaling function
of $T/\sqrt{B}$ proposed by Volovik \cite{Volovik93}. The loss of the entropy deduced from the specific heat data
well above $T_c$ in underdoped cuprates has revealed much important information on the normal state
pseudogap\cite{Loram93,Loram97,Loram00}. 

In this paper we report our experimental data of the electronic specific
heat of $Y_{0.8}Ca_{0.2}Ba_2Cu_3O_{6+x}$ measured from $\sim 7K$ to room temperatures in an external magnetic
field up to $13T$. We choose Calcium doped $YBa_2Cu_3O_{6+x}$ (YBCO) samples because the substitution of $Y^{3+}$
by $Ca^{2+}$ adds holes to the material. This allows physical properties of YBCO-type cuprates to be investigated
in both over- and under-doped regimes.

The data reported in this paper were taken on polycrystalline $Y_{0.8}Ca_{0.2}Ba_2Cu_3O_{6+x}$ samples of about 1g
in  weight each, prepared by conventional solid state reaction. The oxygen content was controlled by the annealing
temperature and determined from the weight loss after annealing, with an uncertainty ~0.01 in x. We measured the
specific heat using a differential technique in which the difference in specific heat between the sample under
investigation and a related reference sample is measured with a resolution of about $1:10^4$ (i.e. $1\%$ of the
electronic contribution)\cite{Loram83}. The majority of the phonon contribution is backed off by the reference
sample and a simple and reliable correction for residual differences in phonon terms allows us to separate
accurately the electronic specific heat $C_{el}$ from the total \cite{Loram93}. In this paper we are mostly
concerned with the change of the specific heat between a finite field and a zero field determined in separate
runs. Whilst this quantity is independent of the phonon specific heat, the inherent reproducibility of the
differential technique coupled with the elimination of the bulk of the phonon term greatly increases the
reliability with which small field dependent changes can be determined.

The rest of the paper is arranged as follows. In Sec. \ref{sec2}, our experimental results are presented and
discussed.  Emphasis is given to the scaling behavior of the specific heat in a finite field and the temperature
dependence of the upper critical field. Sec. \ref{sec3} gives a brief summary.

\section{Results and Discussions}
\label{sec2}

\begin{figure}
\begin{center}
\epsfig{file=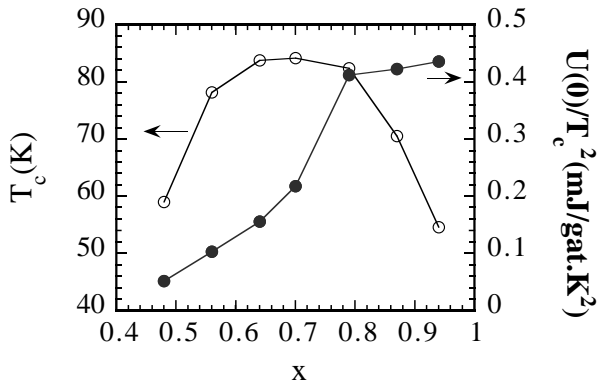, width=8cm,clip=,angle=0}
\end{center}
\caption{Superconducting transition temperature $T_c$ and $U(0)/T_c^2$ for
$Y_{0.8}Ca_{0.2}Ba_2Cu_3O_{6+x}$, where U(0) is the condensation energy at zero temperature. The optimal
doping is located at $x=0.7$ with $T_c=84K$. $U(0)/T_c^2$ is nearly doping independent when
$x>0.79$ but decreases with decreasing doping when $x<0.79$.}
\label{fig1}
\end{figure}

We have measured seven $20\%$ Ca substituted YBCO samples. Zero field measurements on this system have been
discussed previously \cite{Loram97}. Fig.\ref{fig1} shows the critical transition temperature $T_c$ for all the
samples. The optimal doping is located at $x=0.7$ and $T_c=84K$. The condensation energy at zero temperature
$U(0)$, rescaled by $T_c^2$, is also shown in this figure. The value of $U(0)$, as will be discussed later, is
determined from the difference of the free energy density between the normal and superconducting states.
$U(0)/T_c^2$ is nearly doping independent when $x>0.79$ but drops rapidly with decreasing $x$ when $x<0.79$.
$U(0)/T_c^2$ at the optimal doping $x=0.7$ $(T_c=84K)$ is about half of the value at $x=0.79$ $(T_c=80K)$. This
rapid decrease in $U(0)$ is due to the opening of the normal state pseudogap at a planar hole density $p_{crit}
\sim 0.19 holes/CuO_2$\cite{Loram93,Loram97}. The abrupt change in the doping dependence of the condensation
energy around $x=0.79$ might be a manifestation of quantum critical phenomena. Similar changes have also been
found in the doping dependence of the superfluid stiffness\cite{Pana00} as well as many other physical quantities
\cite{Tallon00,Shen00}.

\subsection{Specific heat and the scaling behavior}

The specific heat is an important quantity in characterizing low-energy excitations (see for example ref.
\cite{Junod90,Junod96}). It provides a direct measure of thermal excitations (spectral weight) in the electronic
spectrum. In a Fermi liquid system, the temperature dependence of the electronic specific heat coefficient,
$\gamma (H,T) \equiv C_{el}(H,T)/T$ at low temperatures reflects the energy dependence of the density of states of
low-lying excitations.

\begin{figure}
\begin{center}
\epsfig{file=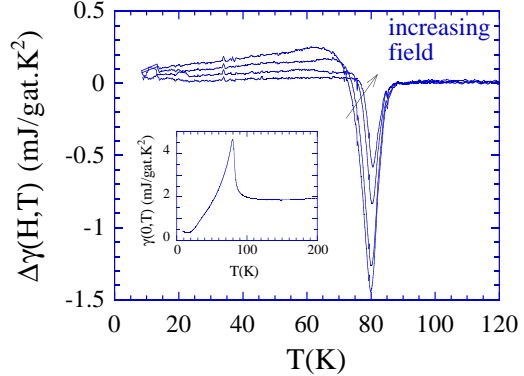,width=7cm,clip=,angle=0}
\end{center}
\caption{Changes of the electronic specific heat coefficient $\Delta\gamma (H,0)$ at four different magnetic
fields for $Y_{0.8}Ca_{0.2}Ba_2Cu_3O_{6.79}$. $H=2T$, $4T$, $9T$, and $13T$ from bottom to top.} \label{fig2}
\end{figure}

Fig. \ref{fig2} shows the change of the specific heat coefficient $\Delta\gamma (H,T) \equiv\gamma (H,T)-\gamma
(0,T)$ in four fields $H=2T$, $4T$, $9T$ and $13T$ for $Y_{0.8}Ca_{0.2}Ba_2Cu_3O_{6.79}$. Since no corrections
from the lattice contribution to the raw data of $\Delta\gamma$ are involved, there is no ambiguity in this
quantity. In the normal state ($T > T_c + 10K$) , $\gamma$ is unchanged by the applied fields within an
experimental uncertainty of $\sim 0.01 mJ/gat.K^2$. At zero field (see the inset of Fig.\ref{fig2}, $\gamma $
shows a sharp peak around $T_c$. This peak is gradually suppressed by increasing field.

In type-II superconductors in an applied magnetic field, there are two types of 
low-lying excitations: vortex core bound states and extended quasiparticle states 
which evolve smoothly into the bulk zero-field quasiparticle states at large distances 
from the vortex \cite{Volovik93}. In a conventional $s$-wave superconductor, 
the low energy density of states is dominated by the core bound states, 
since the extended states are fully gapped and the core level spacing 
is about a factor $\Delta /\varepsilon _{F}$ smaller than the quasiparticle excitation 
gap, where $\Delta $ is the energy gap and $\varepsilon_F$ is the Fermi energy. 
Since the number of vortices scales linearly with $H$, the low energy density of states 
is also proportional to $H$. This means that at low temperatures and when $H\ll H_{c2}$, 
the specific heat scales linearly with $H/H_{C2} $, i.e. $C_{el}(H)\sim H$.

However, in the mixed state of a superconductor with line nodes, Volovik\cite{Volovik93} pointed out that the low energy density of states is dominated by contributions from extended quasiparticle states rather than the core bound states. It was shown that the core bound states contribute less to the entropy than the extended quasiparticles by a factor that diverges logarithmically with the inter-vortex separation. This is due to the effect of Doppler shift caused by supercurrents circulating around a vortex. Since the inter-vortex distance is proportional to $H^{-1/2}$ and the number of vortices is proportional to $H$, it follows that the density of states at the Fermi level is proportional to $\sqrt{H}$ and the specific heat varies as $T\sqrt{H}$ in the limit $T\rightarrow 0$. Furthermore, from the $\sqrt{H}$\ scaling behavior of the excitation spectrum, Simon and Lee \cite{Simon97} showed that low temperature thermal and transport coefficients of a d-wave superconductor should scale with $\sqrt{H}/T$. In particular, they predicted that the specific heat has a simple scaling form
\begin{equation}
C_{el}=T\sqrt{H}F_{C}\left( T/\sqrt{H}\right) , \label{scale}
\end{equation}
where $F_{C}$ is a undetermined scaling function. This scaling prediction holds for clean d-wave superconductors
and for energy scales small compared to the maximum gap scale. In the dirty limit, Kubert and Hirschfeld
\cite{Kubert98} showed that the low-lying excitations are still dominated by the extended states, but the specific
heat behaves as $H\ln H$ at low fields. The $\sqrt{H}T$ term of the specific heat was first identified by Moler et
al \cite{Moler94} for YBCO single crystals and later by other groups \cite{Revaz98,Wright99}. The $\sqrt{H}/T$
scaling behavior has also been reported in the specific heat measurements of YBCO\cite{Junod97,Revaz98,Wright99}.

\begin{figure}
\begin{center}
\epsfig{file=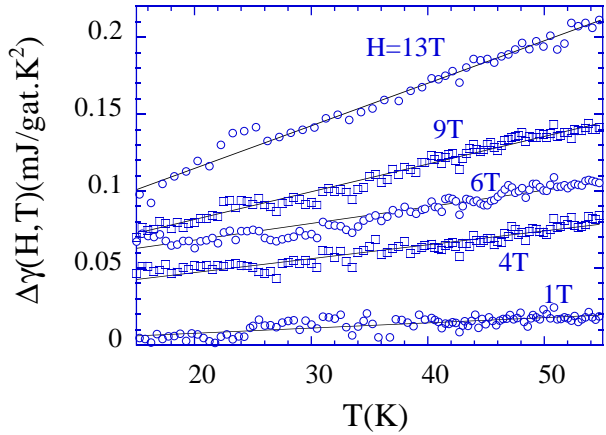,width=8cm,clip=,angle=0}
\end{center}
\caption{Low temperature behaviours of $\Delta\gamma (H,T)$ at five different fields for
$Y_{0.8}Ca_{0.2}Ba_2Cu_3O_{6.79}$. The solid lines are linear fits to the data.}
\label{fig3}
\end{figure}

\begin{figure}
\begin{center}
\epsfig{file=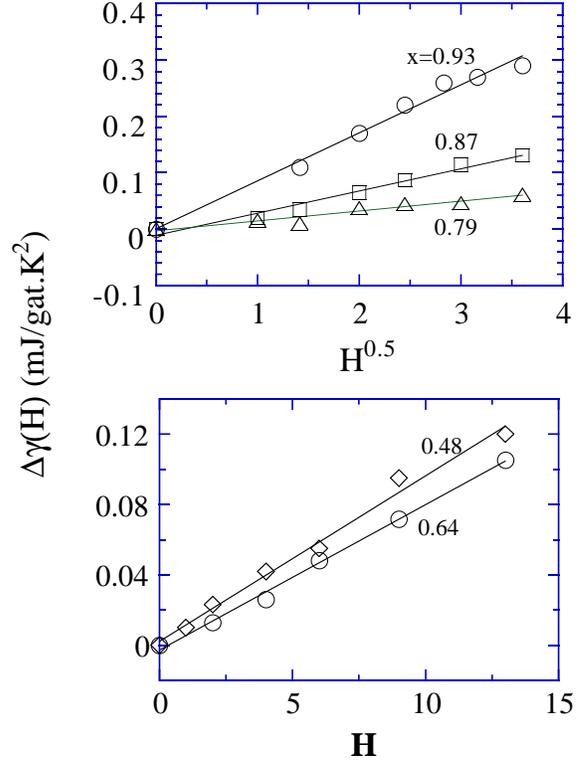,width=8cm,clip=,angle=0}
\end{center}
\caption{The specific heat coefficient $\Delta \gamma (H, 0)$ as a function of $\sqrt{H}$ for the
overdoped $Y_{0.8}Ca_{0.2}Ba_2Cu_3O_{6+x}$ (upper panel) and a function of $H$ for the
underdoped $Y_{0.8}Ca_{0.2}Ba_2Cu_3O_{6+x}$ (lower panel).  }
\label{fig4}
\end{figure}

For Ca doped YBCO, we find that the field dependent $\Delta\gamma $ varies linearly with temperature at low
temperatures (Fig. \ref{fig3}) and can be fitted by the equation
\begin{equation}
\Delta\gamma (H, T) \approx \Delta\gamma (H, 0) + \alpha (H) T
\end{equation}
within experimental errors. The first term is just the coefficient of the INITIAL linear $T$ term of $C_{el}$ and
is predicted to vary as $\sqrt{H}$ in an ideal $d$-wave superconductor. For overdoped
$Y_{0.8}Ca_{0.2}Ba_2Cu_3O_{6+x}$ (the upper panel of Fig. \ref{fig4}), we indeed find that $\Delta\gamma (H,0)
\sim \sqrt{H}$, with slopes comparable with that found for optimally doped single crystal YBCO
\cite{Moler94,Wright99} and the theoretical predication\cite{Volovik93}. However, for underdoped
$Y_{0.8}Ca_{0.2}Ba_2Cu_3O_{6+x}$ (the lower panel of Fig. \ref{fig4}), we find that $\Delta\gamma (H,0)$ varies
linearly with $H$. This is different from the $\sqrt{H}$ scaling behavior of a clean d-wave superconductor and
also from the $H\ln H$ scaling behavior predicted for a dirty d-wave superconductor. We do not know if this
violation of the $\sqrt{H}$ scaling law of $C_{el}$ in these underdoped samples is due to the effect of normal
state pseudogap or just because the temperatures we have measured are not low enough to
probe the characteristic low-energy feature of a $d$-wave superconductor. Further investigation on the field
dependence of the specific heat of underdoped high-$T_c$ materials at low temperatures is desired.

\begin{figure}
\begin{center}
\epsfig{file=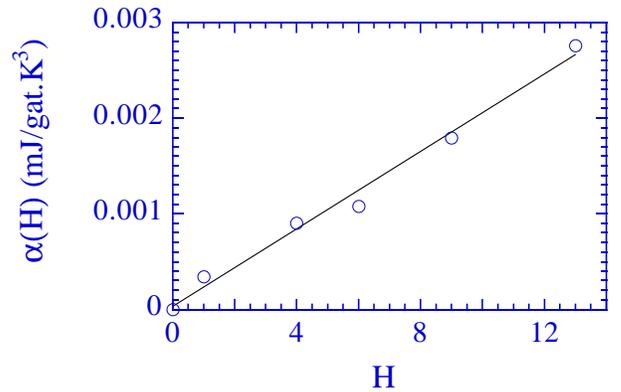,width=8cm,clip=,angle=0}
\end{center}
\caption{ $\alpha (H)$ versus $H$ for $Y_{0.8}Ca_{0.2}Ba_2Cu_3O_{6.79}$. }
\label{fig5}
\end{figure}

The field dependence of $\alpha (H)$ is more complicated and less accurately determined than $\Delta\gamma (H)$.
However, for the $x=0.79$ sample, we find that $\alpha (H)$ varies almost linearly with $H$ (Fig. \ref{fig5}).
This linear $H$ dependence of $\alpha (H)$ is not what one may expect from the scaling relation (\ref{scale}).

\subsection{Entropy}

The electronic entropy $S$ is one of the parameters sensitive to both superconducting condensation and normal
states pseudo-gap and can be determined from the temperature integration of thespecific heat coefficient
\begin{equation}
S=\int_{0}^{T}\gamma (T)dT
\end{equation}

\begin{figure}
\begin{center}
\epsfig{file=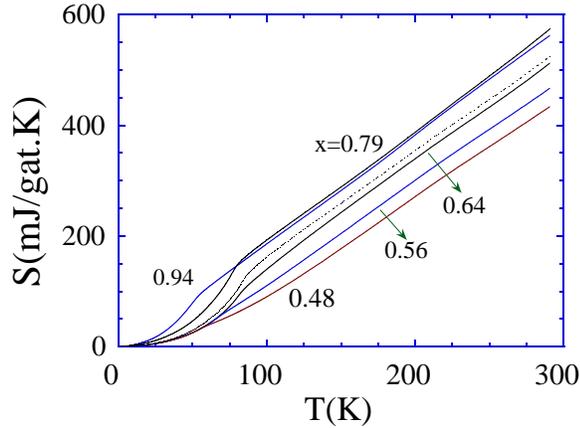,width=8cm,clip=,angle=0}
\end{center}
\caption{Electronic entropy versus temperature for $Y_{0.8}Ca_{0.2}Ba_2Cu_3O_{6+x}$
at zero field. $x$ for the curves from top to bottom are 0.93, 0.87, 0.79, 0.70, 0.64,
0.56, and 0.48, respectively. The dashed line is for the optimal doping (x=0.70).
The entropy loss is observed in normal state in optimal
as well as underdoped samples. }
\label{entropy}
\end{figure}

Our data of $S$ for $20\%$ Ca doped YBCO at zero field are shown in Fig. \ref{entropy}. There is an abrupt change
in the slope of $S$ at $T_c$ due to the specific heat jump. For the overdoped samples (x=0.93, 0.87 and 0.79), the
normal state entropy extrapolates to zero at zero temperature. However, for the optimal and underdoped samples (x=
0.7, 0.64, 0.56 and 0.48)  the normal state entropy extrapolates to a negative value at zero temperature. This
means that some of the entropy is lost in the normal state in these materials. This loss of entropy has been
attributed to the opening of the normal state pseudogap. Similar results have been observed in other cuprate
systems.\cite{Loram93,Loram00}

\begin{figure}
\begin{center}
\epsfig{file=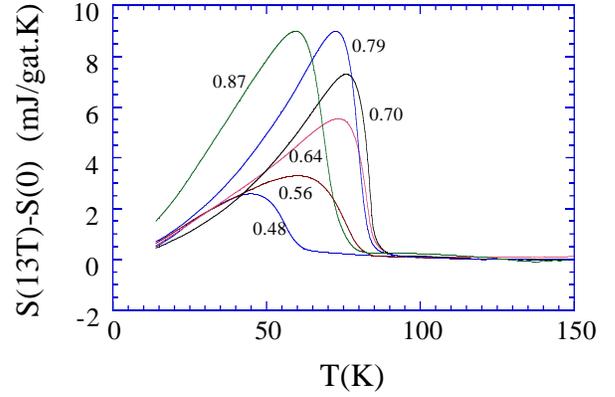,width=8cm,clip=,angle=0}
\end{center}
\caption{Difference of the entropy between zero field and $H=13T$ for
$Y_{0.8}Ca_{0.2}Ba_2Cu_3O_{6+x}$. The value of $x$ is given besides each curve. The
external field affects strongly the condensation energy but has almost no effect on the normal state
entropy. }
\label{fig6}
\end{figure}

Fig. \ref{fig6} shows the enhancement of the entropy at a $13T$ magnetic field for different oxygen contents.  The
superconducting condensation is suppressed by an external field below $T_c$. This causes the increase of the
entropy. In the normal state, the entropy is almost unchanged by the external field. This suggests that the normal
state pseudogap is unaffected by an external field, consistent with the NMR data for underdoped
YBCO\cite{Gorny99}.

\begin{figure}
\begin{center}
\epsfig{file=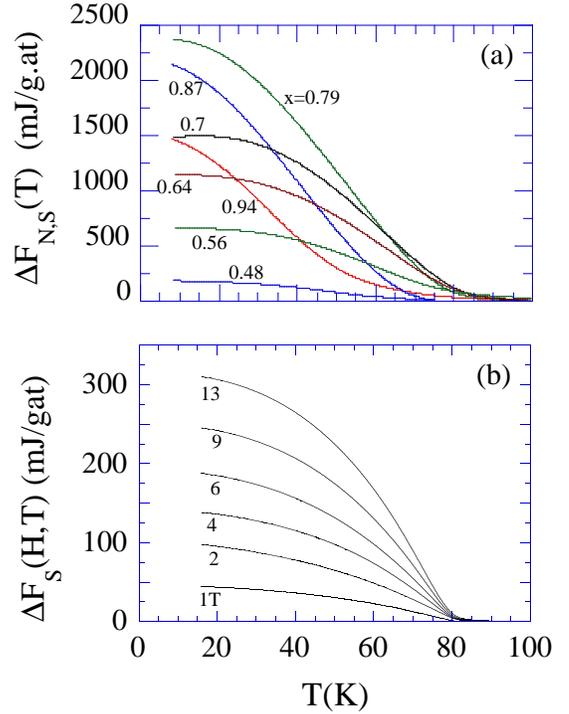,width=8cm,clip=,angle=0}
\end{center}
\caption{(a) $\Delta F_{N.S}(T) = F_N(0,T)-F_S(0,T)$ versus temperature for
$Y_{0.8}Ca_{0.2}Ba_2Cu_3O_{6+x}$. (b) $\Delta F_S(H,T)=$ $F_S(H,T)-F_S(0,T)$ at six different
fields for $x=0.79$.}
\label{fig7}
\end{figure}

\subsection{Free energy}

From the entropy, one can further find out the free energy density. In particular, by integrating out the entropy difference between the normal and superconducting states at zero field, the change of the free energy
\begin{equation}
\Delta F_{N,S}(T) = F_N(0,T)-F_S(0,T) =\frac{H_c^2 (T)}{8\pi}
\end{equation}
can be obtained, where $H_c(T)$ is the thermodynamic field. At zero temperature $\Delta F_{N,S}(0)$ is the
condensation energy $U(0)$. Our results of $\Delta F_{N,S}(T)$ are shown in Fig. \ref{fig7}(a) for different
oxygen contents. The behavior of this quantity is different for overdoped and underdoped samples. With decreasing
temperatures, $\Delta F_{N,S}(T)$ increases more rapidly in the overdoped samples than in the underdoped ones.

The change of the free energy density in a field in the superconducting state $\Delta F_S(H,T) = F_S(H,T) -
F_S(0,T)$  can be similarly obtained. Fig. \ref{fig7}(b) shows $\Delta F_S(H,T)$ for the $x=0.79$ sample. The
largest $\Delta F_S$ at $13T$ and $0K$ is about $220mJ/gat$, much smaller than $U(0)\sim 1550mJ/g.at$. This
indicates that the upper critical field $H_{c2}$ at zero temperature is much larger than our maximum field of 13T.

\subsection{The upper critical field}

In recent years, there has been intensive investigation on the temperature dependence of the upper critical  field
$H_{c2}$ in high-$T_c$ superconductors. MacKenzie et al.\cite{Mac93} measured the field dependence of the in-plane
resistivity of overdoped $Tl_2Ba_2CuO_{6}$ and found that $H_{c2}(T)$ shows a positive curvature at low
temperatures. Similar temperature dependence of $H_{c2}$ has also been found in the transport measurements of
other overdoped materials \cite{Oso93}. To understand this puzzling upward curvature of $H_{c2}(T)$, a number of
theories have been proposed\cite{Ges98,Joynt90,Kotliar96,Ale97}, invoking thermal and quantal fluctuations.
Recently, Wen et al measured temperature dependence of the magnetic moment of overdoped $Bi_2Sr_{2-x}La_xCuO_4$
and $La_{2-x}Sr_xCuO_4$ and found that there are two transitions in the M(T) curve in an applied
field\cite{Wen99,Wen00}. The field dependence of the lower transition (or crossover) temperature indeed exhibits
an upturn at low temperatures, similar to the upper critical field found in the transport measurements. However,
the field dependence of the upper transition temperature behaves very differently. It shows a negative curvature,
same as for the upper critical field in conventional superconductors. They argued that the transition at higher
temperature corresponds to the true upper critical field $H_{c2}$ and the upward curvature of the critical field
at lower temperature is in fact due to the formation of a bulk superconducting phase coherence through Josephson
coupling between some superconducting clusters with a transition temperature higher than the bulk transition
temperature\cite{Ges98}.

We can estimate the upper critical field $H_{c2}$ from the free energy data obtained above if we assume that  the
magnetisation obeys the London expression (see for example Ref \cite{Junod96}). In the mixed state, the difference
of the free energy $\Delta F_S$ in the superconducting state can be deduced from the reversible magnetization M:
\begin{equation}
\Delta F_S(H,T) = -\int_0^H MdH. \label{free}
\end{equation}
The London formula of the magnetization in the mixed states is approximately given by\cite{Tin}:
\begin{equation}
M= - \frac{H_{c1}}{8\pi\ln \kappa }\ln \frac{bH_{c2}}{H},
\label{mag}
\end{equation}
where $\kappa =\lambda /\xi $ is the Ginsberg-Laudau parameter, $\lambda $ is the penetration depth, and $b$ is a
constant. The lower and upper critical fields $H_{c1}$ and $H_{c2}$ are related to the thermodynamic field by the
formula $H_{c1}H_{c2}=H_c^2\ln \kappa$. Substituting (\ref{mag}) into (\ref{free}) and carrying out the
integration, we find that
\begin{equation}
\Delta F_S(H,T) = \frac{H \Delta F_{N,S}(T)}{H_{c2}(T)}\left[ \ln \frac{bH_{c2}(T)}{H} + 1 \right] .
\label{hc2}
\end{equation}

\begin{figure}
\begin{center}
\epsfig{file=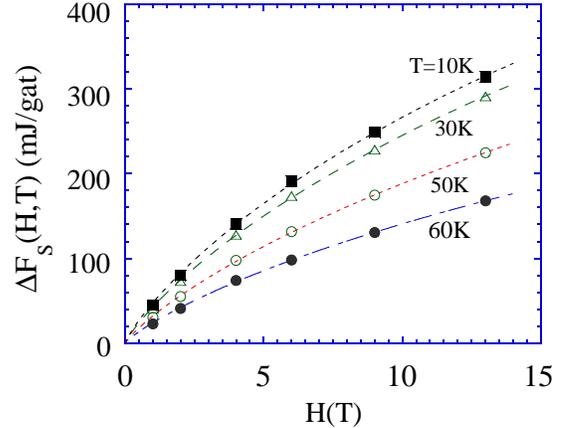,width=8cm,clip=,angle=0}
\end{center}
\caption{$\Delta F_S(H,T)$ versus H at different temperatures for $Y_{0.8}Ca_{0.2}Ba_2Cu_3O_{6.79}$. The dot and
dash  lines are the fits using Eq.(7)} \label{fig8}
\end{figure}
\begin{figure}
\begin{center}
\epsfig{file=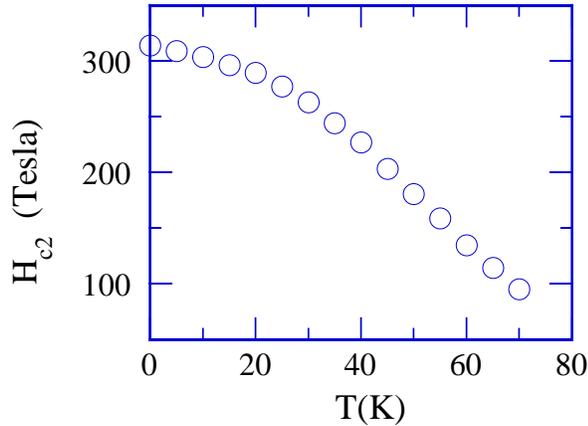,width=8cm,clip=,angle=0}
\end{center}
\caption{The upper critical field $H_{c2}$ for $Y_{0.8}Ca_{0.2}Ba_2Cu_3O_{6.79}$.}
\label{fig9}
\end{figure}

Fig. \ref{fig8} shows $\Delta F_s)$ as a function of $H$ at different temperatures for the $x=0.79$ sample.  The
the dash and dot lines are the fits of $\Delta F_S$ using Eq.(\ref{hc2}) with the value of $F_{N,S}(T)$ shown in
fig.\ref{fig7}. The values of $H_{c2}$ for this material can be obtained from the fits and the result is shown in
Fig. \ref{fig9} for the $x=0.79$ sample (similar results have been obtained for other samples). Same as for
conventional $s$-wave superconductors, we find that $H_{c2} $ has a negative curvature. Close to $T_c$, $H_{c2}$
varies linearly with temperature and $-dH_{c2}/dT\sim 5T/K$. This agrees with the electronic specific heat
measurement for overdoped $Tl_2Ba_2CuO_{6}$ \cite{Radcliffe96}, where $H_{c2}$ is determined by the changes of
$T_c$ in magnetic fields. The magnitude of $H_{c2}$ is much larger than that determined from the transport
measurement, in agreement with the magnetization measurement data of Wen et al\cite{Wen99,Wen00} for overdoped
$Bi_2Sr_{2-x}La_{x}CuO_{4}$ and $La_{2-x}Sr_{x}CuO_{4}$, but in disagreement with the transport measurement data
for overdoped$Tl_2Ba_2CuO_{6}$ \cite{Mac93} and $Bi_2Sr_2CuO_{y}$ \cite{Oso93}. The difference between the
thermodynamic and transport measurements has been attributed by Wen et al to a macroscopic phase separation of
electrons in overdoped materials. Another alternative is that transport measurements are sensitive to flux-flow
resistance at temperatures much lower than the thermodynamic critical temperature.

\section{Summary}
\label{sec3}

In summary, we have measured the electronic specific heat of $Y_{0.8}Ca_{0.2}Ba_2Cu_3O_{6+x}$ using a
differential technique from $7K$ to room temperature in a magnetic field up to 13T. We have discussed the physical
properties of the specific heat, the entropy, the free energy density and the upper critical field. For optimal
and overdoped samples, we found that the field dependent specific heat coefficient scales with $\sqrt{H}$ at low
temperatures, in agreement with other experimental measurements. However, for underdoped samples, we found that
the low temperature specific heat coefficient scales linearly with $H$, inconsistent with the $\sqrt{H}$ scaling
law as expected for clean $d$-wave superconductors. Further investigation on the field dependence of low
temperature specific heat in clean underdoped high-$T_c$ superconductors should be carried out to resolve this
discrepancy. The upper critical field for the overdoped sample ($x=0.79$) shows a downward curvature. This agrees
with the results of the magnetization measurements but differs from those of the transport measurements. It
suggests that the overdoped materials might be very inhomogeneous in an external field and the upward curvature of
$H_{c2}$ observed in the resistivity measurement is probably due to the Josephson coupling between the
superconducting clusters produced by electronic phase separation.
\section{Acknowledgements}

J.L and T.X. acknowledge the hospitality of the Interdisciplinary Research Center in Superconductivity of the
University of Cambridge, where part of the work was done, and acknowledge the financial support from the National
Natural Science Foundation of China and from the special funds for Major State Basic Research Projects of China.

\end{multicols}

\end{document}